\documentclass[default, trackchanges]{aastex701}
\usepackage{afterpage}
\usepackage{placeins}

\begin{document}

\title{The Nature of an Unidentified X-ray Source in the 7 Ms Chandra Deep Field-South Catalog}
\author[orcid=0009-0005-6904-6220, sname='Sullivan']{Quinn P. Sullivan}
\affiliation{Department of Astronomy and Astrophysics, 525 Davey Lab, The Pennsylvania State University, University Park, PA 16802, USA }
\affiliation{Department of Physics, 104 Davey Laboratory, The Pennsylvania State University, University Park, PA 16802, USA}
\email{qps5048@psu.edu}

\author[orcid=0000-0002-6990-9058, sname='Yu']{Zhibo Yu}
\affiliation{Department of Astronomy and Astrophysics, 525 Davey Lab, The Pennsylvania State University, University Park, PA 16802, USA }
\affiliation{Institute for Gravitation and the Cosmos, The Pennsylvania State University, University Park, PA 16802, USA}
\email{zvy5225@psu.edu}

\author[orcid=0000-0002-0167-2453, sname='Brandt ']{William N. Brandt}
\affiliation{Department of Astronomy and Astrophysics, 525 Davey Lab, The Pennsylvania State University, University Park, PA 16802, USA }
\affiliation{Department of Physics, 104 Davey Laboratory, The Pennsylvania State University, University Park, PA 16802, USA}
\affiliation{Institute for Gravitation and the Cosmos, The Pennsylvania State University, University Park, PA 16802, USA}
\email{wnbrandt@gmail.com}

\author[orcid=0000-0002-9036-0063, sname='Luo']{Bin Luo}
\affiliation{School of Astronomy and Space Science, Nanjing University, Nanjing 210093, People’s Republic of China}
\affiliation{Key Laboratory of Modern Astronomy and Astrophysics (Nanjing University), Ministry of Education, People’s Republic of China}
\email{bluo@nju.edu.cn}

\author[orcid=0000-0002-4436-6923, sname='Zou']{Fan Zou}
\affiliation{Department of Astronomy, University of Michigan, 1085 S University, Ann Arbor, MI 48109, USA}
\email{fanzou01@gmail.com}

\begin{abstract}
In the 7 Ms Chandra Deep Field-South catalog, only one source, XID 912, was highly significantly detected in X-rays but had no formally reported counterparts in the UV, optical, infrared, or radio bands. We identified its potential JWST and VLT VIMOS counterparts and measured the corresponding aperture photometry to construct its spectral energy distribution (SED). We fitted this SED using \texttt{CIGALE}. The results indicate that the source is most likely an off-nuclear ultraluminous X-ray source, rather than a background active galactic nucleus.

\end{abstract}

\section{INTRODUCTION} 
\label{sec:intro}
The 7 Ms Chandra Deep Field-South (CDF-S) survey \citep{Luo+2017} is the deepest X-ray survey to date. Our source, XID 912, was the only highly significantly detected source in the 7 Ms CDF-S that had no formal counterpart in any other bands \citep{Luo+2017}. It had an ACIS Extract no-source probability of $P_B \approx 10^{-13}$. XID 912 had 73.2 counts from 0.5--2.0 keV and no detection in the 2.0--7.0 keV band. It is also located 7.5\arcsec\ from the known X-ray emitting galaxy XID 916 at a spectroscopic redshift of $z = 0.105$ \citep{Szokoly+2004}. There were two scenarios for the nature of XID 912: (1) an off-nuclear X-ray source residing $\sim\!15$ kpc away from the center of XID 916, or (2) an unrelated high-redshift dusty active galactic nucleus (AGN) \citep{Luo+2017}. We investigated the nature of XID 912 by leveraging the new sensitive coverage from JWST and VLT. We extracted the spectral energy distribution (SED) and fitted it with \texttt{CIGALE} \citep{Boquien+2019, Yang+2022}.

\section{DATA AND METHODS} 
\label{sec:data}
We inspected images taken by various telescopes released after the 7 Ms CDF-S catalog was published and only selected data for which we could visually identify a potential counterpart near the X-ray position. Eight filters met this requirement: two VLT VIMOS filters (\textit{U} and \textit{R} bands) and six JWST Near-Infrared Camera filters (F115W, F150W, F200W, F277W, F356W, and F444W). Together, these filters provided coverage from 3765 \AA\ to 44393 \AA. 

\begin{figure}[hp]
    \centering
    \begin{minipage}[b]{0.47\textwidth}
        \centering
        \includegraphics[width=\textwidth]{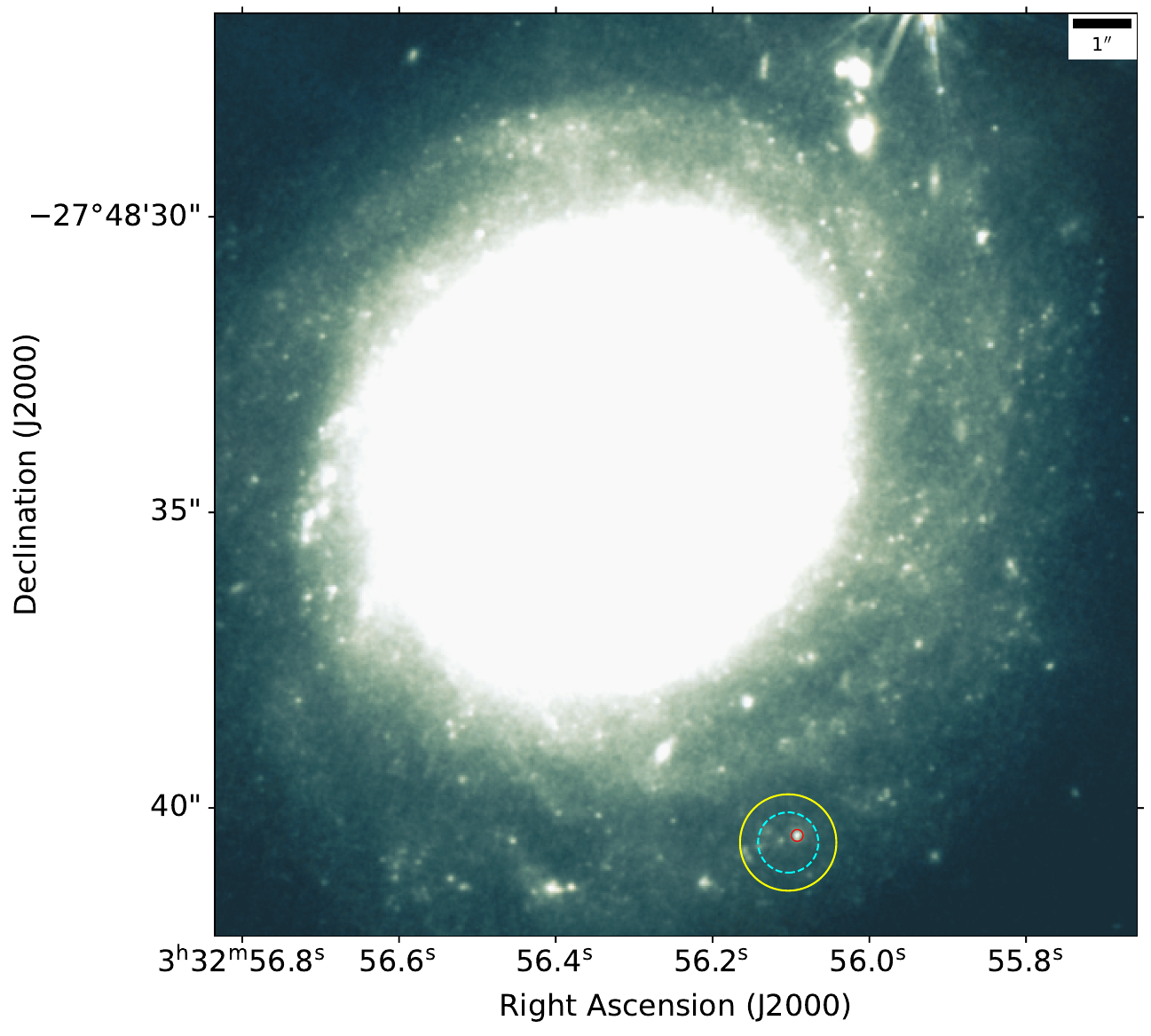}
    \end{minipage}
    \hfill
    \begin{minipage}[b]{0.47\textwidth}
        \centering
        \includegraphics[width=\textwidth]{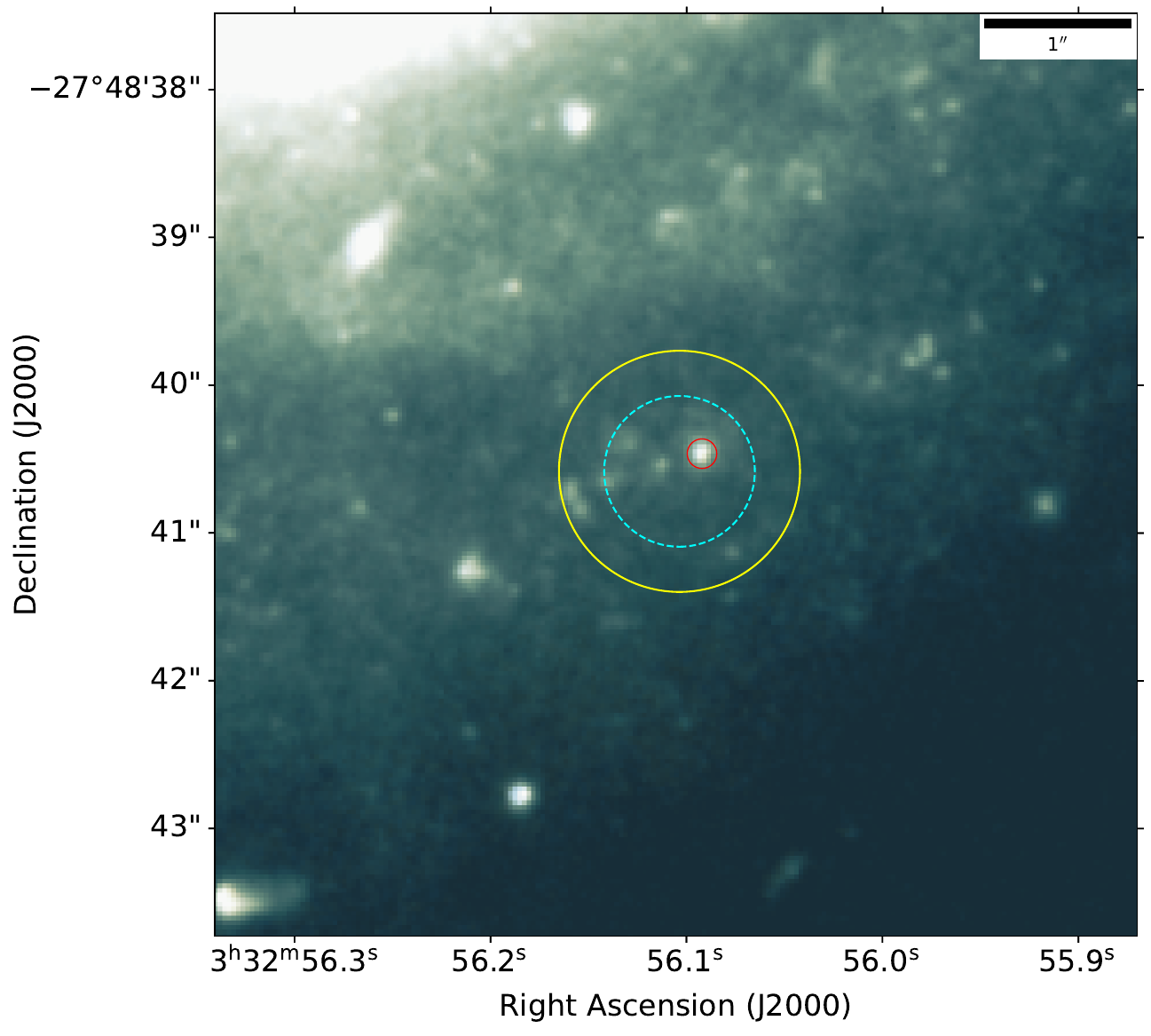}
    \end{minipage}
    
    \vspace{0.5cm}  

    \begin{minipage}[b]{0.47\textwidth}
        \centering
        \includegraphics[width=\textwidth]{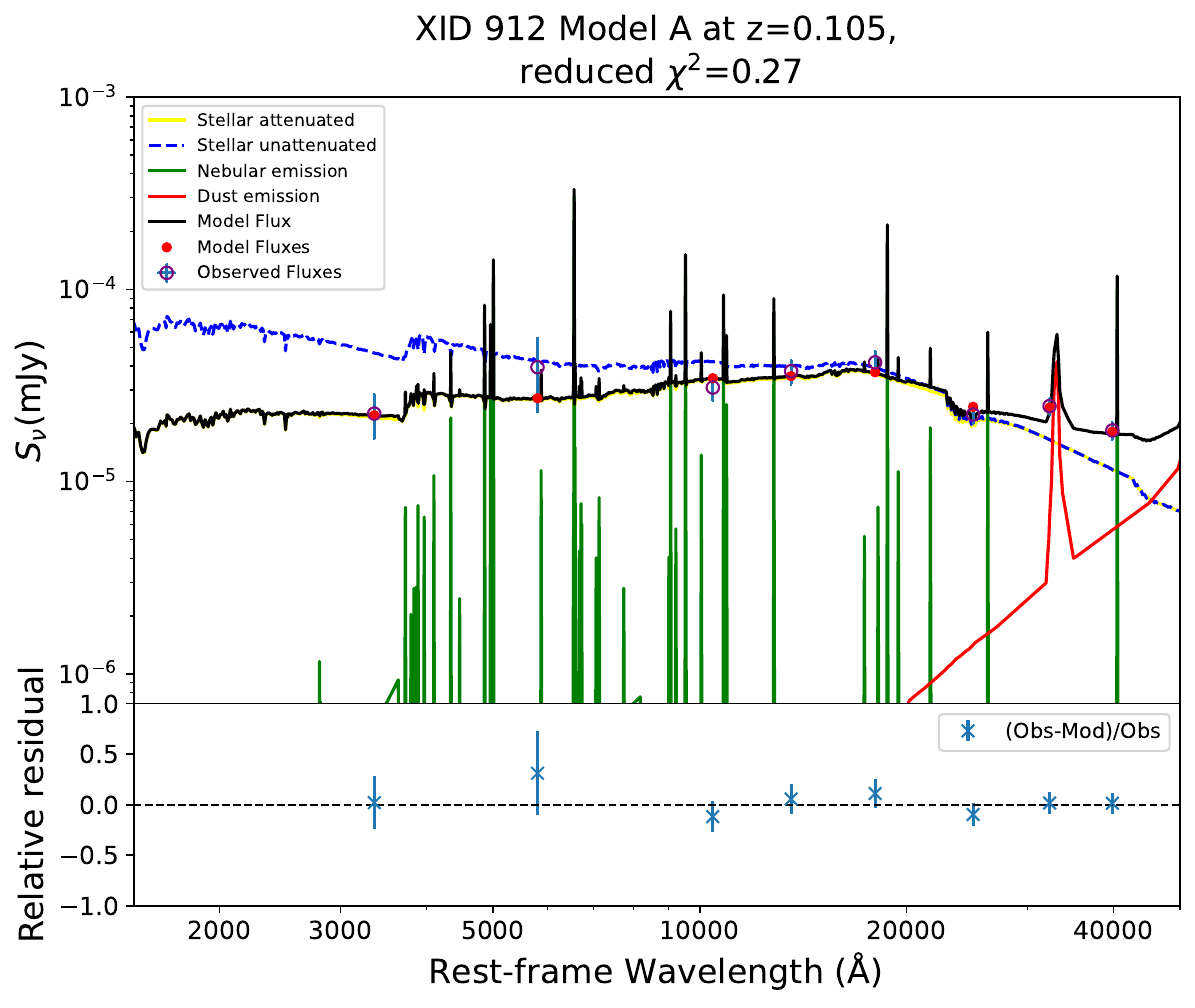}
    \end{minipage}
    \hfill
    \begin{minipage}[b]{0.47\textwidth}
        \centering
        \includegraphics[width=\textwidth]{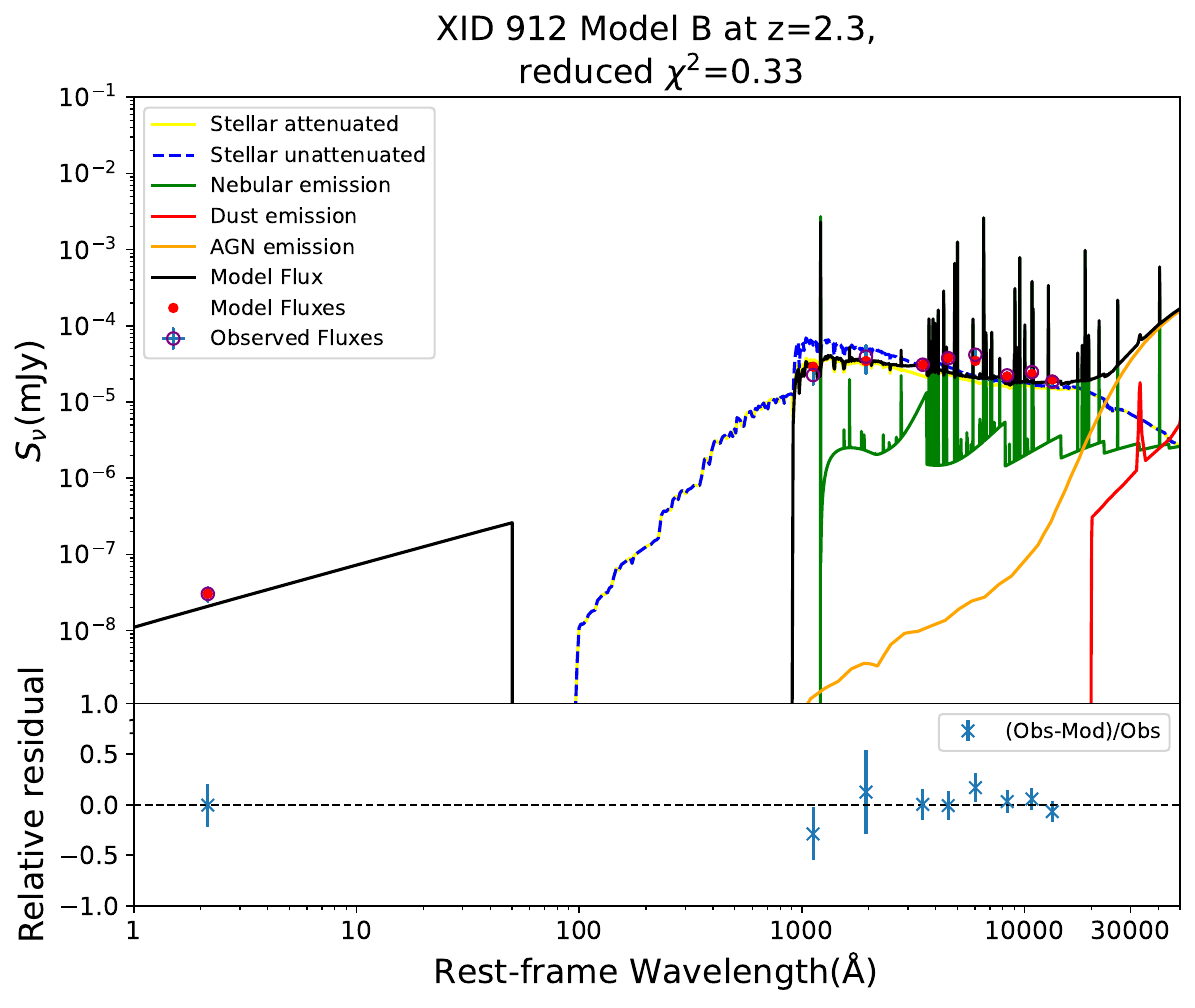}
    \end{minipage}
    \vspace{0.5cm}  

    \caption{\textit{Top-left}: Composite image of XID 912 and its potential host galaxy XID 916 using JWST filters (F150W, F200W, and F277W). The yellow solid circle and cyan dashed circle show the 90\% and the 1$\sigma$  X-ray positional uncertainty of XID 912, respectively \citep{Luo+2017}. We defined the red circle as our photometric aperture for the JWST filters. A 1.0\arcsec\ scale bar is in the upper right. \textit{Top-right}: Zoomed image centered on the reported position of XID 912 from \cite{Luo+2017}. All circle definitions are identical to those in the top-left panel. \textit{Bottom-left}: Best-fit SED using Model A assuming $z = 0.105$. \textit{Bottom-right}: Best-fit SED using Model B. No redshift was assumed.}
    \label{fig:mainfigure}
\end{figure}

We used the JWST F150W image to identify potential counterparts and found five within the 90\% X-ray positional uncertainty (see top panels of Figure \ref{fig:mainfigure}). We considered the brightest JWST source (RA = 53.233717 deg, Dec = \textminus27.811240 deg) as our primary counterpart. We subtracted the JWST simulated point spread functions from our JWST data and found the resulting rms to be similar to the background rms, indicating that our source is point-like. For JWST, we performed aperture photometry using a 0.1\arcsec\ radius aperture. An annulus centered at the same position with an inner radius of 0.4\arcsec\ and an outer radius of 2.0\arcsec\ was used for background subtraction. For VLT, we set the aperture radius to 0.35\arcsec\ in the \textit{R} band and 0.4\arcsec\ in the \textit{U} band. The inner and outer radii of the background annulus were set to 2 and 7 times the aperture radius. Aperture corrections were applied; the corrections were 1.31--1.73 for JWST bands \citep{Rigby+2023} and 1.32 for both VLT filters \citep{Nonino+2009}. We calculated the flux uncertainties using Equation 3 from \cite{Nyland+2017}. The SNRs were 6.5--9.5 in the JWST bands and 2.5 and 4.1 for the VLT \textit{R} band and \textit{U} band, respectively. We used these aperture-corrected fluxes and uncertainties to construct a SED. 

We fitted this SED using \texttt{CIGALE} with two different models. Model A models scenario (1) in Section \ref{sec:intro}. In this scenario, we assume XID 912 has the same redshift as XID 916 and is an X-ray source embedded in a star cluster. The detection in the optical and infrared is mostly contributed by the star cluster. Model A adopts similar settings to \cite{Turner+2021} (bottom-left panel of Figure \ref{fig:mainfigure}). It uses the Star Formation History (SFH) module \texttt{sfh2exp} with a very short e-folding time to model the instantaneous burst expected for a star-forming region and the dust emission module \texttt{dale2014} \citep{Dale+2014}. Model B describes our scenario (2) and adopts similar settings to the AGN model in \cite{Zou+2022}. We simultaneously fitted photometric redshifts over the range $z = 0.0$ to $z = 6.0$ with a step size of 0.05 (bottom-right panel of Figure \ref{fig:mainfigure}). 

\section{RESULTS AND DISCUSSION} 
\label{sec:results}
Model A and Model B provided similarly good fits, so we cannot rule out either scenario based on the reduced-$\chi^2$. For Model B, the redshift probability distribution function strongly favored $1.5<z<3.0$ and had a prominent peak at $z = 2.3$. At $z = 2.3$, the Chandra 0.5--2.0 keV band samples X-rays at \mbox{1.7--6.6 keV} in the rest-frame, indicating that the source of the X-ray emission is the power-law continuum. In this band, the soft X-ray spectrum of XID 912 has a very steep effective power-law photon index of $\Gamma_{\text{eff}} = 3.1 \pm 0.5$ \citep{Luo+2017}. The lower limit of this $\Gamma_{\text{eff}}$ is even higher than the sampled range in \cite{Liu+2021} and is a 2.7$\sigma$ outlier compared to the mean. Also, the JWST images reveal that XID 912 is on a disk spiral arm of XID 916. For these reasons, we believe that XID 912 is likely an off-nuclear X-ray source residing in XID 916 at $z = 0.105$.

Assuming this scenario, we calculated the rest-frame 2--10 keV X-ray luminosity to be $(1.04  \pm 0.19) \times 10^{39}$ \( \mathrm{erg \, s^{-1}} \) using Equation 1 from \cite{Zou+2022} with the 0.5--2.0 keV X-ray flux and $\Gamma_{\text{eff}}$ from \cite{Luo+2017}. This X-ray luminosity should be considered a lower bound, as no intrinsic obscuration was considered. XID 912 would be classified as an ultraluminous X-ray source (ULX). Along with $\Gamma_{\text{eff}} \approx3.1$, XID 912 could be further classified as an ultraluminous supersoft source (ULS) \citep{Urquhart+2016}.  ULSs have a blackbody component that varies on both short-term and long-term timescales with a temperature of $kT_{\mathrm{bb}} \approx 50-150\,\mathrm{eV}$ and little emission above 1 keV. ULSs may represent highly obscured ULXs, with their soft blackbody spectra and variability explained by an expanding and contracting photosphere viewed from a nearly edge-on orientation \citep{Urquhart+2016, King+2023}. For XID 912, the X-ray photons were relatively evenly distributed across the 7 Ms observations, with some low-amplitude variability present \citep{Luo+2017}. Our limited photon statistics prevent us from fitting a blackbody to any single epoch, but fitting a blackbody to the stacked X-ray spectrum (accounting for Galactic extinction) yields a blackbody temperature of $kT_{\mathrm{bb}} = 250^{+150}_{-90} \mathrm{eV}$, which is generally consistent with a ULS. 

We used the same procedure as in Section \ref{sec:data} to perform aperture photometry with only JWST data for the four other bright sources within the X-ray positional uncertainty. These four sources have similar SEDs, so we fitted the SED only for the brightest among them. The results are similar to those for our primary counterpart, with Model B preferring an AGN at $z\approx1.8\text{--}3.2$. Due to the large $\Gamma_{\text{eff}}$ and the fact that these counterparts are on a disk arm, we still prefer the ULX scenario.
 
\section*{Acknowledgments}
The JWST data presented in this article were obtained from the Mikulski Archive for Space Telescopes (MAST) at the Space Telescope Science Institute. The specific observations analyzed can be accessed via \dataset[doi:10.17909/70qh-te67]{http://dx.doi.org/10.17909/70qh-te67}. This paper employs a list of Chandra datasets, obtained by the Chandra X-ray Observatory, contained in the Chandra Data Collection (CDC) at \dataset[doi:10.25574/cdc.437]{https://doi.org/10.25574/cdc.437}.

\FloatBarrier

\bibliography{912}
\bibliographystyle{aasjournalv7}



\end{document}